\documentclass[a4paper,twocolumn,%tightenlines,
english,aps,prb,floatfix,showpacs]{revtex4}
\usepackage[T1]{fontenc}
\usepackage[latin1]{inputenc}
\usepackage{amsmath}
\usepackage{graphics}
\usepackage{amssymb}
\usepackage{verbatim}

\begin{document}
 
\title{Properties of the multicritical point of $\pm J$ Ising spin glasses on the square 
lattice}

\author {Jean C. \surname{Lessa}}

\email{jean@if.ufrj.br}

\affiliation{Departamento de F\'{i}sica, Universidade Estadual de Feira de 
Santana, Campus Universit\'ario, 44031-460
Feira de Santana BA, Brazil}

\author {S.L.A. \surname{de Queiroz}}

\email{sldq@if.ufrj.br}

\affiliation{Instituto de F\'\i sica, Universidade Federal do
Rio de Janeiro, Caixa Postal 68528, 21941-972
Rio de Janeiro RJ, Brazil}

\date{\today}

\begin{abstract}
We use numerical transfer-matrix methods to investigate properties of the multicritical
point of binary Ising spin glasses on a square lattice, whose location we assume
to be given exactly by a conjecture advanced by Nishimori and Nemoto.
We calculate the two largest Lyapunov exponents, as well as linear and non-linear
zero-field uniform susceptibilities, on strip of widths $4 \leq L \leq 16$ sites,
from which we estimate the conformal anomaly $c$, the 
decay-of-correlations exponent $\eta$, and the linear and non-linear 
susceptibility exponents $\gamma/\nu$ and  $\gamma^{nl}/\nu$, with the help of 
finite-size scaling and conformal invariance concepts. Our results are: $c=0.46(1)$;
$0.187 \lesssim \eta \lesssim 0.196$; $\gamma/\nu=1.797(5)$; $\gamma^{nl}/\nu=5.59(2)$.
%%%%%%%%%%%%%%% Ref .1 q.2 Ref. 2, q.2 %%%%%%%%%%%%%%%%%%%%%%%%%%%%%%%%%%%%
A direct evaluation of correlation functions on the strip geometry, and of the 
statistics of the zeroth moment of the associated probability distribution, gives 
$\eta=0.194(1)$, consistent with the calculation via Lyapunov exponents.
%%%%%%%%%%%%%%%%%%%%%%%%%%%%%%%%%%%%%%%%%%%%%%%%%%%%%%%%%%%%%%%%%%%%%%%%%%%%%%%%%
Overall, these values tend to be inconsistent with the universality class of 
percolation, though by small amounts.
The scaling relation $\gamma^{nl}/\nu=2\,\gamma/\nu+d$ (with space dimensionality $d=2$)
is obeyed to rather good accuracy, thus showing no evidence of multiscaling behavior of
the susceptibilities.
\end{abstract}
\maketitle
%\tightenlines
\section{INTRODUCTION}
\label{intro}
The critical behavior of Ising spin glasses  
has been the subject of intensive investigation in the recent past~\cite{nish01}. A 
number of results have been derived, both analytically and numerically; however,
many aspects of interest have not been fully elucidated so far. 
Among these are the properties of the multicritical point which is known to exist 
for suitably low concentrations of antiferromagnetic interactions, even for
two-dimensional lattices in which a spin glass phase is not expected
to occur at non-zero temperatures.  

Here we consider binary ($\pm J$) spin glasses, i.e., we assume that Ising spin-$1/2$
magnetic moments interact via nearest-neighbor couplings $J_{ij}$ of equal strength, and
whose signs are given by the quenched probability distribution:
\begin{equation}
P(J_{ij})= p\,\delta (J_{ij}-J_0)+ (1-p)\,\delta (J_{ij}+J_0)\,.
\label{eq:1}
\end{equation}
In this case, the phase diagram on the temperature-concentration ($T - p$) plane
exhibits a critical line which, for low concentrations $1-p$ of antiferromagnetic bonds, 
separates ferro- and paramagnetic phases~\cite{nish01,prl87_047201}. 
As $p$  decreases, so does the transition temperature. Below a critical concentration
$p_{c}$, ferromagnetic order disappears, and a spin glass phase emerges at $T=0$.
For space dimensionality $d\geq 3$, the spin glass phase extends to finite temperatures 
as well. The {\it Nishimori line} (NL) is a special line on the  $T - p$ plane,
defined by:
\begin{equation}
e^{-2 J_0/T} = \frac{1-p}{p}\qquad\qquad {\rm (NL,}\ p> \frac{1}{2})\ .
\label{eq:2}
\end{equation}
On this line, several exact results have been obtained~\cite{nish01}. In particular, the 
configurationally averaged internal energy is an analytical function of $T$, even at the 
multicritical point (the {\em Nishimori point}, NP) where the NL crosses the 
ferro-paramagnetic phase boundary~\cite{prl61_625}. Furthermore, the NL is invariant 
under renormalization-group transformations, so the NP corresponds to a fixed point.
Numerical work in two-dimensional systems~\cite{jpa35_8171} shows that internal-energy 
fluctuations along the  NL go through a maximum at the NP, thus indicating that the 
latter indeed marks a change in the behavior of the distribution of frustrated
plaquettes. This, in turn, is consistent with the picture that the phase transition 
at the NP  is of geometry-induced nature~\cite{jpa35_8171} (though not necessarily 
in the same universality class of random percolation).

Recently it was predicted~\cite{jpsp71_1198,jpa36_9799} that, on a square
lattice, the NP should belong to a subspace of the $T-p$ plane which is invariant under
certain duality transformations. For $\pm J$ Ising systems
the invariant subspace is given by~\cite{jpsp71_1198,jpa36_9799}
\begin{equation}
p\,\log_2 (1+e^{-2J_0/T})+(1-p)\,\log_2 (1+e^{2J_0/T})= \frac{1}{2}\  .
\label{eq:3}
\end{equation}
The intersection of Eqs.~(\ref{eq:2}) and~(\ref{eq:3}) gives the conjectured exact
location of the NP, namely $p=0.889972 \cdots\,$,  $T/J_0 =0.956729 \cdots\,$, to be 
referred to as CNP.
In previous work~\cite{prl87_047201,prb60_6740,pre65_026113},
approximate estimates for the location of the NP were used in the calculation
of the associated critical exponents, with the overall conclusion that
the transition there does not belong to the universality class of random percolation.
A numerical study of correlation-function statistics at the CNP~\cite{dqrbs03}
points to a similar picture.

Extensions of the conjecture to triangular and honeycomb lattices have been 
proposed~\cite{jpa38_3751,jpsp75_034004}, and verified by numerical studies to
a fairly good degree of accuracy~\cite{prb73_064410}. Evidence thus far
available indicates that the critical properties of the NP in two-dimensional
$\pm J$ Ising systems are universal in the expected (i.e., lattice-independent)
sense~\cite{prb73_064410}, though they belong to a distinct universality class 
from that of percolation.  

Here we use numerical transfer-matrix methods, together with
finite-size scaling and conformal invariance concepts, to investigate critical 
properties of the NP of $\pm J$ Ising spin glasses, on long strips of a square lattice.
We shall assume the location of the multicritical point to be that of the CNP given 
above. Indeed, previous work (see Ref.~\onlinecite{dqrbs03} and references therein)
strongly indicates that, even though the conjecture may turn out not to be exact,
it is certainly a very good approximation to the actual position of the NP.

In Section~\ref{sec:2} we evaluate the central charge, or conformal 
anomaly~\cite{prl56_742}.
As this is given by the coefficient of the finite-size correction to a bulk
quantity (the critical free energy) which is itself not known exactly for the
present case, one has many sources of uncertainty to contend with, not to mention
those intrinsic to the sampling of quenched disorder configurations.  
%%%%%%%%%%%%%%%%%% Ref. 1, q.1 %%%%%%%%%%%%%%%%%%%%%%%%
By working at the CNP, we attempt to eliminate one such source which
is the location of the critical point. In the absence of further exact results, 
whether or not such choice in 
fact introduces systematic distortions can only be found by comparative 
analysis of numerical data pertaining to the problem.   
%%%%%%%%%%%%%%%%%%%%%%%%%%%%%%%%%%%%%%%%%%%%%%%%%%%%%%%%%%%%%%%%%%%%%%%%%%%
In Section~\ref{sec:3} 
%%%%%%%%%%%%%%%%%% Ref. 1, Q.2 and Ref. 2, Q.2 (cont'd) %%%%%%%%%%%%%%%%%%%%%%%%
we calculate the decay-of-correlations exponent related to the zeroth moment
of the correlation-function probability distribution, both via the difference
between the two largest Lyapunov exponents, and by direct evaluation of 
%%%%%%%%%%%%%%%%%%%%%%%%%%%%%%%%%%%%%%%%%%%%%%%%%%%%%%%%%%%%%%%%%%%%%%%%%%%%%%%
correlation functions as done in earlier 
work~\cite{prl87_047201,prb60_6740,dqrbs03}. This, together with the use
of pertinent conformal-invariance relationships, yields further independent evidence
related to the universality properties of correlations at the NP. 
In Section~\ref{sec:4}, both linear and non-linear zero-field susceptibilities are 
investigated. While the former have been evaluated previously 
for square~\cite{prb60_6740}, as well as triangular and honeycomb~\cite{prb73_064410},
lattices, no results for the latter appear to be available. As explained below, the
scaling of non-linear susceptibilities may give indications of multiscaling behavior.
Finally in Section~\ref{sec:conc}, concluding remarks are made.
 
\section{Free energy and central charge}
\label{sec:2}

We consider strips of width $L$ sites and periodic boundary conditions across.
For consistency with earlier work~\cite{prb60_6740}, we used  only even widths,
in order to accommodate possibly occurring unfrustrated antiferromagnetic
ground states (though later results showed in practice that, at least
for the relatively low concentrations of antiferromagnetic bonds around the NP,
no noticeable distortions arise when odd values of $L$ are considered as 
well~\cite{prl87_047201,prb73_064410}).
Appropriate sampling  of quenched disorder
is produced by using strip lengths $N \gg 1$, along which bond configurations
are drawn from the distribution Eq.~(\ref{eq:1}). 
The configurationally-averaged (negative) free energy $f_{L}$ (in units of $k_{B}T$)
is given by
\begin{equation}
 f_{L}=L^{-1}\,\Lambda_{0}(L)\,,
\label{eq:2.1}
\end{equation} 
where $\Lambda_{0}$ is the largest Lyapunov characteristic exponent~\cite{prmsp}, 
extracted from the product of $N \to \infty$ transfer matrices $T_{j}$
which connect site columns  $j$ and $j+1$, i.e.,
\begin{equation}
\Lambda_{0}=\lim_{N\rightarrow \infty}\frac{1}{N}\ln \left\Vert \left( \prod_{j=1}^{N}T_{j}
\right)\vert v_{0}\rangle \right\Vert\,,
\label{eq:2.2}
\end{equation} 
where $ \vert v_{0}\rangle $ is an arbitrary initial vector of unit modulus.
Higher-order exponents may be obtained through iteration of a set of initial
vectors $\vert v_{i}\rangle $, orthogonal both mutually and to  $\vert v_{0}\rangle$,
with adequate reorthogonalization every few steps, to avoid contamination~\cite{prmsp}.

In our calculations we have used  $N\sim 10^{5}\,-\,10^{6}$. The uncertainty
related to the finite number of terms in Eq.~(\ref{eq:2.2}) is estimated as
follows. In order to avoid transient effects, the first $N_{0}\sim 10^3$
iterations are discarded. Accumulated averages are evaluated and stored, 
for each $10^3$ subsequent iterations. From this set of averages, one  then
calculates global averages and their corresponding fluctuations.
In our study, we have always made use of a canonical distribution of
disorder, that is, the  $+J_0$ and $-J_0$ couplings are randomly extracted from a 
reservoir which initially contains exactly as many of each as given by 
Eq.~(\ref{eq:1}), with the value of $p$ corresponding, e.g., to the CNP.
This way, sample-to-sample fluctuations are considerably reduced.   
Final estimates of the free energy and other quantities of interest have been
extracted from arithmetic averages for distinct disorder realizations.

The conformal anomaly, or central charge $c$, which characterizes the universality
class of a conformally-invariant model at the critical point, can be
evaluated via the finite-size scaling of the free energy on a strip with periodic
boundary conditions across~\cite{prl56_742},
\begin{equation}
 f(T_{c},L)=f(T_{c},\infty)+\frac{\pi c}{6L^{2}}+{\cal O}\left(\frac{1}{L^4}\right)\ , 
\label{eq:2.3}
\end{equation} 
where $f(T_{c},\infty)=\lim_{L\rightarrow \infty}\,f(T_{c},L)\,$
is a regular term which corresponds to the bulk system free energy.
For disordered systems, Eq.~(\ref{eq:2.3}) is expected to hold, with
the (configurationally-averaged) free energy given by Eq.~(\ref{eq:2.1}), and
$c$ taking the meaning of an \emph{effective} conformal anomaly~\cite{npb285_687}.

\begin{figure}
%{\centering \resizebox*{3.6in}{!}{\includegraphics*{nishcc-1.eps}}}
{\centering \resizebox*{3.4in}{!}{\includegraphics*{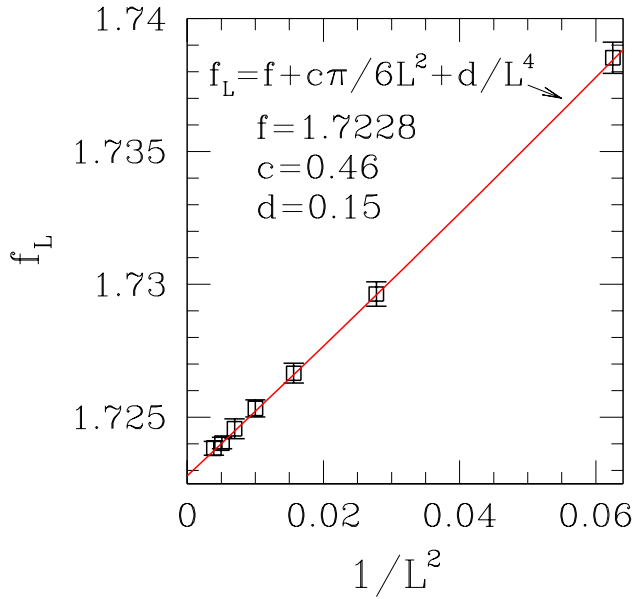}}}
\caption{(Color online) Negative free energy at the conjectured Nishimori point, for 
strip widths $4 \leq L \leq 16$ and $N=10^{6}$, against $1/L^{2}$. The line
uses the central estimates from a parabolic least-squares fit to data, via 
Eq.~(\protect{\ref{eq:2.3}}), which 
gives $f_L=1.7228(1)+0.46(1)\pi/6L^2+0.15(9)/L^4$.} 
\label{fig:fe}
\end{figure}

For our estimates of the effective central charge, we set $T$ and $p$ corresponding 
to the CNP, and took averages of the free energy  $f(T_{c},L)$ over three independent
realizations. Figure~\ref{fig:fe} shows the free energy at the CNP for $4 \leq L \leq 
16$, against $1/L^2$. A linear least-squares fit of the data gives $c=0.478(4)$. 
This is close to, but some $3\%$ off, the result given in
Ref.~\onlinecite{prl87_047201}, $c=0.464(4)$, which presumably was taken at those 
authors' own estimate of the location of the NP, $p=0.8906(2)$.
On the other hand, the above result is compatible
with the value corresponding to percolation in the Ising model, namely
$c_p=5\sqrt{3}\,\ln2/4\pi\approx 0.4777$~\cite{npb515_701}.

Incorporating curvature via the $L^{-4}$ correction, as suggested by Eq.~(\ref{eq:2.3}),
results in the same $f(T_{c},\infty)$ (to within $0.01\%$) as in the linear
extrapolation. However, the conformal anomaly estimate is changed to $c=0.46(1)$,
which encompasses the result quoted in Ref.~\onlinecite{prl87_047201} but appears
incompatible with Ising-model percolation.  Earlier work in pure~\cite{dq00} and
unfrustrated random-bond~\cite{npb515_701,dq95} Ising systems indicates that, for a 
square lattice, the $L^{-4}$ term provides a important contribution towards stability and
accuracy of free-energy extrapolations (note, however, that here the curvature effect is 
imperceptible to the naked eye, see Figure~\ref{fig:fe}). 
Should the same trend hold in the present case, $c=0.46(1)$ would appear to be the most 
reliable of the two estimates produced here.

\section{The exponent $\eta$}
\label{sec:3}
As a consequence of the preservation of conformal invariance at a second-order phase
transition, the correlation length $\xi_{L}$ on a strip geometry with periodic boundary 
conditions across (calculated at the critical point of the corresponding bulk 
two-dimensional system) is  connected to the decay-of-correlations exponent $\eta$, by 
the relationship~\cite{jpa17_L385}
\begin{equation}
\xi_{L}=\frac{L}{\pi\,\eta}\ .
\label{eq:3.1}
\end{equation}

For strips of homogeneous spin systems, the inverse of the dominant correlation length 
(related to the  slowest-decaying critical correlations) is given by the first 
spectral gap of the transfer matrix~\cite{nig90}. A straightforward adaptation for 
disordered cases can be devised through the replacement of pure-system eigenvalues by 
their counterparts in a disordered environment, namely the Lyapunov characteristic
exponents~\cite{prmsp,npb515_701,nig90}. 
One can then calculate the correlation length $\xi_L$ 
(and thus the exponent $\eta$ from  Eq.~(\ref{eq:3.1})), via
\begin{equation}
 \xi_{L}^{-1}=\Lambda_{L}^{(0)}-\Lambda_{L}^{(1)},
\label{eq:3.2}
\end{equation} 
where $\Lambda_{L}^{(0)}$, $\Lambda_{L}^{(1)}$ are, respectively, the largest and 
second-largest Lyapunov exponents. 

In the present case, the dominant correlations are ferromagnetic, and the 
Hamiltonian is invariant under global spin inversion. Therefore, in order to
calculate $\Lambda_{L}^{(0)}$  ($\Lambda_{L}^{(1)}$), it is sufficient to iterate  
$ \vert v_{0}\rangle $ ($ \vert v_{1}\rangle $) which is even (odd) under that
same symmetry~\cite{nig90,glaus87}, with no need for decontamination of the iterates
of $ \vert v_{1}\rangle $.

%%%%%%%%%%%%%%%%%%%%%%%%%%%%%%% Ref. 1, Q.2; Ref. 2, Q.2 (cont'd)%%%%%%%%%%%%%%%%
Before going further, one must recall that correlation functions at the NP 
are multifractal~\cite{prl87_047201,pre65_026113,dqrbs03,prb73_064410,mc02a,mc02b}.
In other words, the rate of decay (against distance $R$) of the moments of assorted 
orders, $G_n (R)$,  of the correlation-function distribution is regulated by 
a set of exponents $\{\eta_n\}$, which are not connected by a single
gap exponent, as is the case for pure systems where $\eta_n=n\,\eta_1$.
The exponent estimated via Eqs.~(\ref{eq:3.1}) and~(\ref{eq:3.2}) is in fact $\eta_0$,
which characterizes the zeroth--order moment of the correlation-function 
distribution, i.e. it gives the typical, or most probable, value of this 
quantity (see, e.g., Ref.~\onlinecite{dq97} and references therein). One has, in the 
bulk,
\begin{equation}
G_0 (R) \equiv \exp\left[ \ln \langle  \sigma_{0}\sigma_{R} \rangle \right]_{\rm 
av}\sim R^{-\eta_0}\ .
\label{eq:3.3}
\end{equation}

%%%%%%%%%%%%%%%%%%%%%%%%%%%%%%%%%%%%%%%%%%%%%%%%%%%%%%%%%%%%%%%%%%%%%%%%%%%%
By evaluating estimates of $\eta_0$ for the range of strip widths within reach
of our computational facilities, we can, in principle, extrapolate the sequence 
$L/\pi\xi_{L}$  to $L\rightarrow \infty$, this way presumably accounting for higher-order 
finite-size corrections to  Eq.~(\ref{eq:3.1}). Earlier results for pure 
systems~\cite{dq00} again indicate that $L^{-2}$ is a convenient variable against
which to set up an extrapolation scheme.

%%%%%%%%%%%%%%%%%%%%%%%%%% Ref. 1, Q.2; Ref. 2, Q.2 (cont'd)%%%%%%%%%%%%%%%%%%%%%%
We can also calculate correlation functions directly on a strip, as done in 
Refs.~\onlinecite{prl87_047201,dqrbs03,prb73_064410}, and examine the
behavior of their zeroth-order moment against distance, from which the appropriate
correlation length can be extracted and plugged back into Eq.~(\ref{eq:3.1}).
Note that negative values of the correlation function will be present upon
sampling; this is not an unsurmountable obstacle for the calculation of logarithmic 
averages here, as it is known that the distribution at the NP is sharply peaked 
close to unity~\cite{dqrbs03}. Consequently, one can deal instead with 
absolute values, as can seen by recalling that a logarithmic average is the same 
as the logarithm of a geometric mean: as long as the overall sign of the product
of all terms is positive (which we have reason to believe here), it does not
matter that some (few) are negative.   
%%%%%%%%%%%%%%%%%%%%%%%%%%%%%%%%%%%%%%%%%%%%%%%%%%%%%%%%%%%%%%%%%%%%%%%%%%%%%%%%%
\begin{figure}
%{\centering \resizebox*{3.8in}{!}{\includegraphics*{nisheta-1.eps}}}
{\centering \resizebox*{3.4in}{!}{\includegraphics*{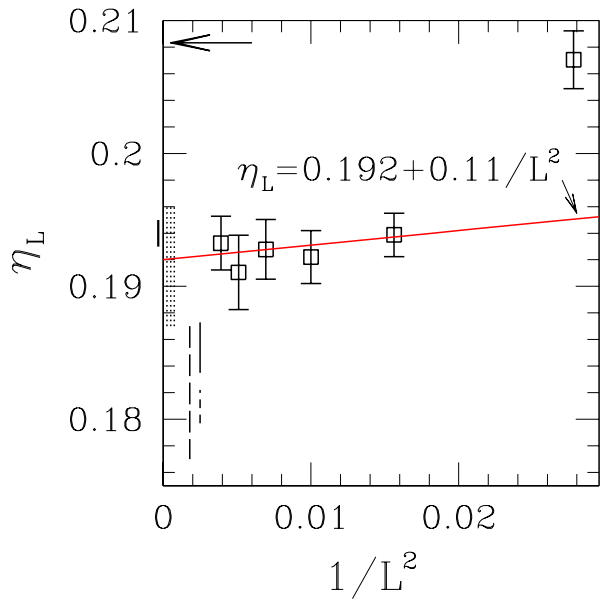}}}
\caption{(Color online) Exponent $\eta_{L}$ against 
$1/L^{2}$. Data are for $L=6, \cdots ,16$. The arrow pointing to the vertical axis 
indicates the percolation value,
$\eta_p=5/24$. The straight line is a least-squares fit of $L=8-16$ data.
Shaded area close to the vertical axis indicates rough limits of confidence for 
extrapolation of finite-size data. 
%%%%%%%%%%%%%%%%%%%%%%%%%% Ref. 1, Q.2; Ref. 2, Q.2 (cont'd)%%%%%%%%%%%%%%%%%%%%%%
Vertical bar to the left of vertical axis gives range of $\eta_0$ from direct
evaluation of zeroth moment of correlation-function distribution (see text).
%%%%%%%%%%%%%%%%%%%%%%%%%%%%%%%%%%%%%%%%%%%%%%%%%%%%%%%%%%%%%%%%%%%%%%%%%%%%%%%%%%
Vertical lines at left of graph show ranges of some 
recent estimates of $\eta_1$. Full line, Refs.~\protect{\onlinecite{prl87_047201}},
\protect{\onlinecite{dqrbs03}}; short dashes, Ref.~\protect{\onlinecite{prb73_064410}};
long dashes, Ref.~\protect{\onlinecite{prb60_6740}}.
} 
\label{fig:eta}
\end{figure}

In Figure~\ref{fig:eta}, we present $\eta_{L}$ calculated via Eqs.~(\ref{eq:3.1}) 
and~(\ref{eq:3.2}), against $1/L^{2}$. On increasing $L$ from $4$ (not shown)
to $8$, there is a decreasing trend in $\eta_L$ which appears to halt, and turn
to a roughly $L-$independent behavior,  when values corresponding to larger $L \leq 16$ 
are considered. Given the circumstances, the safest course of action is (i) to assume 
that the approximately constant behavior will not change significantly for larger $L$
outside our computational capability range, and (ii) to extrapolate the data at hand
in as simple a manner as possible, treating the results with a large dose of skepticism.

From a linear least-squares fit of $L=8-16$ data against $L^{-2}$, shown in 
Figure~\ref{fig:eta}, we get a central extrapolated estimate $\eta_0 \approx 0.192$.
By considering the error bars associated to finite-size estimates, it appears
that any value in the range $0.187 \lesssim \eta_0 \lesssim 0.196$ (the extent of 
the shaded area in the Figure) would be plausible. 

%%%%%%%%%%%%%%%%%%%%%%%%%% Ref. 1, Q.2; Ref. 2, Q.2 (cont'd)%%%%%%%%%%%%%%%%%%%%%%
We also evaluated correlation functions directly, and estimated the zeroth moment  
of their probability distribution. Following Ref.~\onlinecite{dqrbs03},
we used $L=10$, and strip length $N=10^7$ columns; we found that the best
set of results was for correlations calculated along the strip, for distances
$1\leq x \leq 18$; when plotted on a semi-logarithmic scale, our data
show slight curvature for $1 \leq x \leq 3$, and set in to a very good straight 
line  for $4\leq x \leq 18$, from which one gets $\eta_0=0.194(1)$ (shown in the 
Figure, as a thick bar immediately to the left of the vertical axis) via 
Eq.~(\ref{eq:3.1}) . This is consistent with, and  more 
accurate than, the extrapolation of the Lyapunov-exponent data given above.
%%%%%%%%%%%%%%%%%%%%%%%%%%%%%%%%%%%%%%%%%%%%%%%%%%%%%%%%%%%%%%%%%%%%%%%%%%%%%

Comparison against data from previous work is as follows.
Numerical estimates from direct evaluation of the first moment of the probability
distribution of spin-spin correlation functions give 
$\eta_1=0.182(5)$ (square lattice, approximate location of the NP
at $p=0.8905(5)$)~\cite{prb60_6740}; $\eta_1=0.1854(19)$ 
(square lattice, approximate location of the NP
at $p=0.8906(2)$)~\cite{prl87_047201}; $\eta_1=0.1854(17)$ (square lattice, 
CNP)~\cite{dqrbs03};  
$\eta_1=0.181(1)$ (triangular and honeycomb lattices)~\cite{prb73_064410}.
All are displayed in Figure~\ref{fig:eta}, for ease of visualization. 
While overlap between these and the error bars of the present result is 
not better than marginal, it is clear that all estimates, for $\eta_0$ and $\eta_1$,
exclude the percolation value, $\eta_{p}=5/24=0.208333\cdots$~\cite{stauffer94}
(shown in the Figure by an arrow) by a safe gap.

We sum up the situation as follows.
Using Eq.~(\ref{eq:3.2}) as a definition of the correlation length for random
systems is well justified in theory~\cite{prmsp,npb515_701,nig90}, 
%%%%%%%%%%%%%%%%%%%%%%%%%%% Ref. 1 Q.2, Ref. 2, q.2 (cont'd)%%%%%%%%%%%%%%%5
and gives the inverse decay rate of the zeroth moment of the correlation-function 
distribution. As seen above, in the present case the associated exponent $\eta_0$ 
appears to differ slightly from $\eta_1$ which relates to the first moment. This is 
probably the rule rather than the exception; indeed,
%%%%%%%%%%%%%%%%%%%%%%%%%%%%%%%%%%%%%%%%%%%%%%%%%%%%%%%%%%%%%%%%%%%
it has been shown
that, for unfrustrated Ising systems, the finite-size scaling of 
numerical estimates of $\eta_0$ derived in the context of  
Eqs.~(\ref{eq:3.1}) and~(\ref{eq:3.2}) differs
from that of results obtained directly from the spatial decay of correlation 
functions~\cite{dq95,dq97}. 
Though in that case the origin of the discrepancy was traced to effects of
the marginal disorder operator~\cite{dq97} known to arise in the absence of frustration,
the analogous operator structure at the NP is not known so far. However, it seems
plausible to ascribe the small differences between the same two groups of results
here, to similar causes.  

\section{Susceptibilities}
\label{sec:4}
The uniform zero-field magnetic susceptibility  $\chi_{L}$ on a strip is given by
the second derivative of the free energy, relative to a uniform field $h$: 
\begin{equation}
 \chi_{L}=\left[ \dfrac{\partial^{2} f_{L}}{\partial h^{2}}\right] _{h=0}\,.
\label{eq:4.1}
\end{equation} 
As usual in the numerical calculation of derivatives, care must be taken to avoid
introduction of spurious errors. We have considered an infinitesimal field
$\delta h=10^{-4}$ (in units of $J_0$), for the finite differences used in the
differentiation denoted in Eq.~(\ref{eq:4.1}). We have also used the same configuration 
of bonds 
(that is, the same sequence of pseudorandom numbers) for the comparison of free energies
at different field values: free energies of the same bond geometry have to be
subtracted. Thus, fluctuations in the finite differences used in the calculation of 
derivatives are much smaller than those for the free energies 
themselves~\cite{prb60_6740}. For the calculations reported in this Section,
we typically used strip lengths $N=10^7$.

Finite-size scaling arguments suggest the following behavior for $\chi_{L}$  at the
critical point $T_c$:
\begin{equation}
\chi_{L}\sim L^{\gamma/\nu}\,,
\label{eq:4.2}
\end{equation} 
where $\gamma$ and $\nu$ are, respectively, the exponents characterizing the 
singularities of bulk uniform susceptibility and correlation length. 

Another quantity of interest is the non-linear
susceptibility $\chi_{L}^{(nl)}$, given in terms of the power-law expansion of the
magnetization $m$:
\begin{equation}
 m=\chi\,h-\chi^{(nl)}\,h^{3}+\cdots\,,
\label{eq:4.3}
\end{equation} 
where
\begin{equation}
\chi^{(nl)}\equiv\,\dfrac{\partial^{3} m}{\partial h^{3}}=\left[ \dfrac{\partial^{4} f}
{\partial h^{4}}\right] _{h=0}.
\label{eq:4.4}
\end{equation}
The nonlinear susceptibility at criticality obeys a finite-size scaling relationship
similar to Eq.~(\ref{eq:4.2}), with the replacement~\cite{bd85} $\gamma  \rightarrow  
\gamma^{nl}$. This quantity has been investigated in the context of critical phenomena
in both pure~\cite{bd85,yuri97} and (quantum) spin-glass magnets~\cite{gbh94}.
The numerical procedures described above, for the calculation of derivatives, 
are followed here as well. 

\begin{figure}
%{\centering \resizebox*{3.8in}{!}{\includegraphics*{nishchi-1.eps}}}
{\centering \resizebox*{3.4in}{!}{\includegraphics*{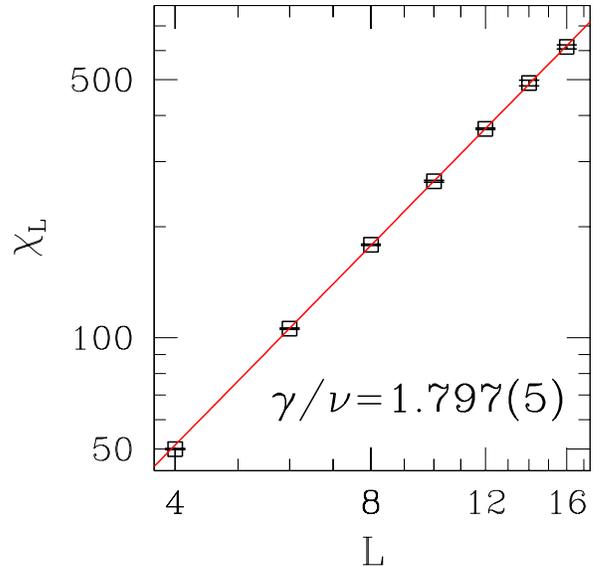}}}
\caption{(Color online) Double-logarithmic plot of zero-field linear susceptibility  
$\chi_{L}$ at the conjectured Nishimori point, for  $L=4-16$. The line is a single-power 
law least-squares fit to $L=6-16$ data, 
enabling the estimation of $\gamma/\nu$ with use of Eq.~(\protect{\ref{eq:4.2}}).} 
\label{fig:chi}
\end{figure}

In Figure~\ref{fig:chi} we show data for the linear susceptibility $\chi_{L}$,
evaluated at the CNP, which have been fitted to the single power-law form, 
Eq.~(\ref{eq:4.2}). We noticed that the $\chi^2$ per degree of freedom decreases from 
$2.3$, for  a fit including $L=4$ data (from which the estimate $\gamma/\nu=1.82(1)$ is 
extracted), to $0.38$ for a fit of $L=6-16$ data only, 
thus pointing to a clear 
improvement in the quality of fit. The latter procedure gives 
$\gamma/\nu=1.797(5)$, which, for the reasons just mentioned, we assume to be
the best result to be extracted from the present data set. This is just 
consistent, at the margin, with the percolation value
$(\gamma/\nu)_{p}=43/24\approx 1.7917$~\cite{stauffer94}. 

In previous work, the following results have been found:
$\gamma/\nu=1.80(2)$~\cite{prb60_6740} (square lattice, approximate location of the NP 
at $p=0.8905(5)$), $\gamma/\nu=1.795(20)$~\cite{prb73_064410} (triangular lattice). 
Both are consistent with the present estimate.
%\begin{comment}
%\end{comment}

\begin{figure}
%{\centering \resizebox*{3.8in}{!}{\includegraphics*{nishchinl-1.eps}}}
{\centering \resizebox*{3.4in}{!}{\includegraphics*{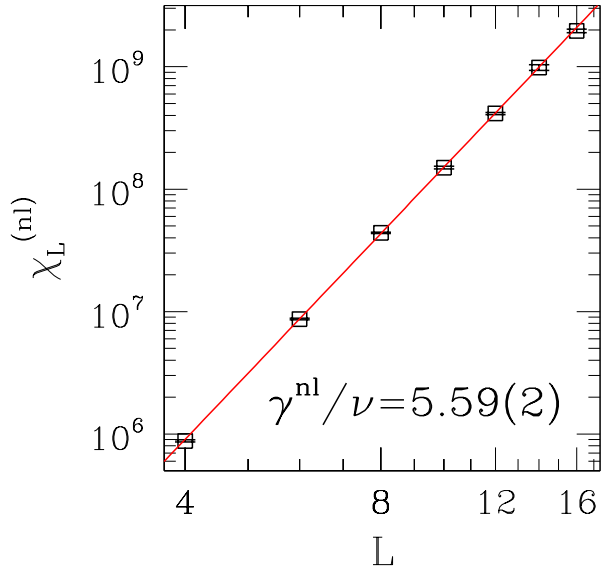}}}
\caption{(Color online) Double-logarithmic plot of zero-field nonlinear susceptibility 
$\chi_{L}^{nl}$ at the conjectured Nishimori point. The curve is a single-power law
least-squares fit to $L=4-16$ data,
enabling the estimation of $\gamma^{nl}/\nu$ with use of Eq.~(\protect{\ref{eq:4.2}}).}
\label{fig:chinl}
\end{figure}

Non-linear zero field susceptibility data ($\chi_{L}^{nl}$), evaluated at the 
CNP, are exhibited in Figure~\ref{fig:chinl}. A fitting procedure, similar to that used 
to extract the linear susceptibility exponent, leads to $\gamma^{nl}/\nu=5.59(2)$
when data for $L=4-16$ are used, and to $\gamma^{nl}/\nu=5.55(2)$ when $L=4$ data
are discarded. However, the  $\chi^2$ per degree of freedom decreases only from
$1.3$ to $0.75$ between the former and latter fits. Bearing in mind that we
are dealing with a small number of finite-size estimates, such a variation does not
warrant discarding $L=4$ data on grounds of a significant improvement in the quality of 
fit. The value $\gamma^{nl}/\nu=5.59(2)$ is consistent with the scaling relation
$\gamma^{nl}/\nu=2\,\gamma/\nu+d = 5.59(1)$ 
(using $\gamma/\nu$ obtained above, and $d=2$ for the space dimensionality). 

\section{Conclusions}
\label{sec:conc}
We have calculated assorted critical quantities at the conjectured exact location
of the Nishimori point (CNP) for square-lattice $\pm J$ Ising spin glasses, namely
$p=0.889972 \cdots\,$,  $T/J_0 =0.956729 \cdots\,$. By working at this fixed location,
we 
%%%%%%%%%%%%%%%%%%%%%%%%%%%%%Ref. 2, Q.1 %%%%%%%%%%%%%%%%%%%%%%%%%%%%
attempt to
%%%%%%%%%%%%%%%%%%%%%%%%%%%%%%%%%%%%%%%%%%%%%%%%%%%%%%%%%%%%%
eliminate one among many sources of uncertainty with which one has to deal
in the study of disordered systems. Of course, whether or not such choice in fact 
introduces systematic distortions can only be found by comparison of a body
of results pertaining to the problem under scrutiny.   

Our extrapolation of finite-size free-energy data, in order to produce an estimate of 
the effective central charge, has been careful by accounting for curvature effects
which are known to be relevant in such circumstances~\cite{npb515_701,dq95,dq00}. 
Our final estimate, $c=0.46(1)$, is consistent with
an earlier result~\cite{prl87_047201}, $c=0.464(4)$, calculated at $p=0.8906(2)$,
and appears to exclude the percolation value~\cite{npb515_701}, $c_p \approx 0.4777$.  
Had we not included curvature effects, we would have reached $c=0.478(4)$ which
would lead to the opposite conclusion.  

%%%%%%%%%%%%%%%%%%Ref. 1, Q.1 %%%%%%%%%%%%%%%%%%%%%%%%%%%%%%%%%%%
We conclude that whatever differences may exist between free energies
evaluated at the CNP, and those calculated at nearby locations such as that
given in Ref.~\onlinecite{prl87_047201}, their effect upon subsequent
estimates of the central charge is not detectable amidst the noise associated to 
other sources of uncertainty. Prominent among these is the current upper 
limit on strip widths, $L \approx 20$, imposed by practical considerations.
%%%%%%%%%%%%%%%%%%%%%%%%%%%%%%%%%%%%%%%%%%%%%%%%%%%%%%%%%%%%%%%%%%

%%%%%%%%%%%%%%%%%%%%%%%% Ref. 1, Q.2; Ref. 2, Q.2(cont'd)%%%%%%%%%%%%%%%%%%%%%%%%%%
We have evaluated finite-size correlation lengths via the difference between the two
largest Lyapunov exponents. With the help of conformal-invariance concepts, these
were used to produce a sequence of estimates of the decay-of-correlations exponent 
$\eta_0$, related to the decay of the zeroth moment of the correlation-function
probability distribution. Though 
such a sequence does not behave as smoothly as its free-energy
counterpart, it seems safe to state that it points to $0.187 \lesssim \eta \lesssim
0.196$. We have also directly calculated correlation functions on a strip, thus
assessing the statistics of the above-mentioned zeroth moment. The corresponding 
result, $\eta_0=0.194(1)$ is consistent with, and more accurate than, that derived from
the Lyapunov exponents. Both estimates slightly differ  
from $\eta_1$, related to the first moment of the same distribution,
for which available estimates~\cite{prb60_6740,prl87_047201,dqrbs03,prb73_064410}
fall in the range $0.180 \lesssim \eta \lesssim 0.187$. In all cases, for 
$\eta_0$ as well as $\eta_1$, the random-percolation 
value~\cite{stauffer94} $\eta_p=0.208333 \cdots$, is definitely excluded.
%%%%%%%%%%%%%%%%%%%%%%%%%%%%%%%%%%%%%%%%%%%%%%%%%%%%%%%%%%%%%%%%%%%%%%%%%%%%%%

From zero-field susceptibility data we obtain $\gamma/\nu=1.797(5)$,
which falls within the range of previous results~\cite{prb60_6740,prb73_064410},
and just about touches the percolation value
$(\gamma/\nu)_{p}=43/24\approx 1.7917$~\cite{stauffer94}, at the lower
end of the error bar. 
%%%%%%%%%%%%%%%%%%%%%% ref. 1, q.2; Ref. 2 , q.2 (ommitted)%%%%%%%%%%%%%%%%%%%
Similarly to the
case discussed in  Ref.~\onlinecite{prb73_064410}, it appears that from $\gamma/\nu$ 
alone it is 
hard to get conclusive evidence, either for or against the behavior at the NP being
in the percolation universality class.

As regards non-linear susceptibilities, our study has been motivated by the well-known
manifestations of multiscaling behavior of correlation functions at
the NP~\cite{prl87_047201,pre65_026113,dqrbs03,prb73_064410,mc02a,mc02b}.
%%%%%%%%%%%%%%%%%%%% Ref.1, Q.3 %%%%%%%%%%%%%%%%%%%%%%%%%%%%%%%%%%%%%%%%%%%%%%%%%
The connection between linear susceptibility $\chi$ and the first moment of
the correlation-function distribution is given through the
fluctuation-dissipation theorem, which (upon invoking standard scaling
arguments~\cite{stan71}) implies the scaling relation $\gamma/\nu=2
-\eta_1$. The non-linear susceptibility $\chi^{(nl)}$, on the other hand, can be 
expressed in terms
of four-point correlations and products of two-point ones~\cite{bd85}. Thus it is not 
obvious {\em a priori} whether any of the multiscaling properties, observed for
the assorted moments of the two-point function, will influence $\chi^{(nl)}$.
Whatever guidance we have on the subject is given by the standard finite-size scaling
relation between the exponents associated to $\chi$ and $\chi^{(nl)}$, 
namely $\gamma^{nl}/\nu=2\,\gamma/\nu+d$. This is established upon
consideration of finite-size scaling properties of the free energy~\cite{bd85},
therefore bypassing any explicit connection to correlation functions.  
Should  multiscaling behavior of magnetization-like quantities occur (via
their connections to aggregated correlation functions), one would expect to see
something similar to the non-constant gap exponents observed for correlation function
statistics  (i.e., non-constant magnetization gap exponents~\cite{stan71}), 
which  would imply breakdown of the relationship just mentioned.
%%%%%%%%%%%%%%%%%%%%%%%%%%%%%%%%%%%%%%%%%%%%%%%%%%%%%%%%%%%%%%%%%%%%%%%%%%%%%%%%%%% 
As seen above, we have found that the relationship
is in fact obeyed, to very good numerical accuracy. Thus, no evidence has been
detected for multiscaling behavior of magnetic susceptibilities.   

%%%%%%%%%%%%%%%%%%%%%%%%%%%%%%%Ref. 1 Q.1 %%%%%%%%%%%%%%%%%%%%%%%%%%%%%%%%%%%%%%%%%%%%%%%%%
NOTE ADDED. After initial submission of this paper, new work came up~\cite{php06}
in which the results of Ref.~\onlinecite{prl87_047201} are extended and reanalysed.
The estimates of the location of the NP, and of the central charge, remain
unchanged at $p=0.8906(2)$ and $c=0.464(4)$, respectively.
%%%%%%%%%%%%%%%%%%%%%%%%%%%%%%%%%%%%%%%%%%%%%%%%%%%%%%%%%%%%%%%%%%%%%%%%%%%%%%%%%%%%%%%%%%%

\begin{acknowledgments}
J.C.L. thanks the Brazilian agency CAPES for partial financial support,
Departamento de F\'\i sica de S\'olidos, UFRJ, for making its research
infrastructure available, and Universidade Estadual de Feira de Santana.    
The research of S.L.A.d.Q. was partially supported by
the Brazilian agencies CNPq (Grant No. 30.0003/2003-0), 
FAPERJ (Grant No. E26--152.195/2002), FUJB-UFRJ, and Instituto do Mil\^enio 
de Nanoci\^encias--CNPq.

\end{acknowledgments}

\end{document}